\documentclass{PoS}

\usepackage{graphicx}
\usepackage{amssymb}
\usepackage{amsmath}
\usepackage{slashed}
\usepackage{caption}
\usepackage{xspace}
\newcommand{\GeV}{\ensuremath{\,\mathrm{GeV}}\xspace}
\newcommand{\nLO}{\nbar \text{LO}\xspace}
\newcommand{\nNLO}{\nbar \text{NLO}\xspace}
\newcommand{\nbar}{\ensuremath{\bar{n}}}
\def\waa{{$W^\pm\gamma\gamma$}}

\title{%
\vspace*{-5cm}
\begin{flushright}%
{\scriptsize %
DCPT/13/100, FTUV-13-0807, IFIC/13-39, IPPP/13/50, KA-TP-15-2013, LPN13-043, SFB/CPP-13-45}%
\end{flushright}
\vspace*{5.5cm}
Di-boson and Tri-boson production at the LHC}

\ShortTitle{Di-boson and Tri-boson production at the LHC}

\author{\speaker{Francisco Campanario}\\
Theory Division, IFIC, University of Valencia-CSIC, E-46980
  Paterna, Valencia, Spain.
        \\
        E-mail: \email{francisco.campanario@ific.uv.es}}

\author{ Christoph Englert\\
       SUPA, School of Physics and Astronomy, University of Glasgow, Glasgow, G12 8QQ, United Kingdom \\
        E-mail: \email{christoph.englert@glasgow.ac.uk}}

\author{ Michael Rauch\\
     Institut f\"ur Theoretische Physik, 
    Universit\"at Karlsruhe, KIT,\\ 76128 Karlsruhe, 
    Germany \\
        E-mail: \email{michael.rauch@kit.edu}}

\author{Sebastian Sapeta\\
     Institute for Particle Physics Phenomenology, Department of
  Physics, Durham University, Durham, United Kingdom \\
        E-mail: \email{sebastian.sapeta@durham.ac.uk}}

\author{Dieter Zeppenfeld\\
     Institut f\"ur Theoretische Physik, 
    Universit\"at Karlsruhe, KIT,\\ 76128 Karlsruhe, 
    Germany \\
        E-mail: \email{dieter.zeppenfeld@kit.edu}}

\abstract{The status of di-boson and tri-boson production is
  shortly review.  Using the VBFNLO and the LOOPSIM  package, approximated
  results at NNLO QCD are given for WZ production. Results for \waa ~+ jet at
  NLO QCD are also shown.}

\FullConference{XXI International Workshop on Deep-Inelastic Scattering and Related Subject -DIS2013,\\
		22-26 April 2013\\
		Marseilles,France}

\begin{document}
\section{Introduction}
Di-boson and tri-boson production processes provide a rich source of
information at hadron colliders and have been already intensively studied both
from the theoretical and the experimental side.  They are important
backgrounds to both Standard Model~(SM) and beyond standard model~(BSM)
searches. As a signal, they yield information on triple and quartic gauge
couplings.

In this proceedings article, a short overview of the theoretical status of di-boson
production and recent
results beyond NLO QCD for $WZ$ production will be presented in
Section~\ref{sect:diboson}. A brief overview of tri-boson production and results for $W 
\gamma \gamma  +jet $ will be given in Section~\ref{sect:triboson}

\section{Di-boson Production}
\label{sect:diboson}
NLO QCD corrections for di-boson production at hadron colliders were computed
in Refs.~\cite{dibosonQCD}. The size of these corrections, for inclusive
cuts, ranges in the order of the 40$\%$ to 100$\%$. The big impact of the NLO
corrections is mainly due to the appearance of new sub-processes at this order as part of the real corrections, namely, 
gluon initiated processes. These are enhanced at the LHC due to the
large gluonic pdfs and partially compensate the suppression on $\alpha_s$. 

\begin{figure}[h!]
\begin{minipage}{1\columnwidth} 
  \includegraphics[width=0.25\columnwidth]{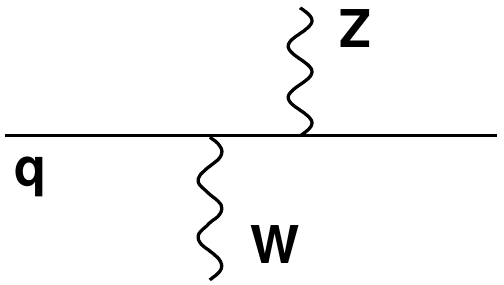} \hfill
  \includegraphics[width=0.25\columnwidth]{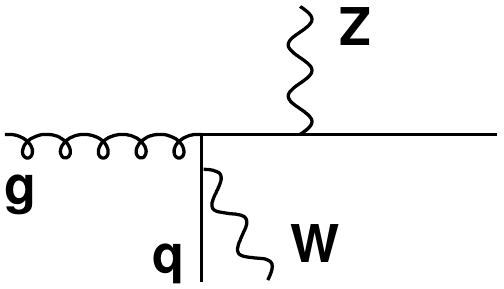}\hfill
  \includegraphics[width=0.25\columnwidth]{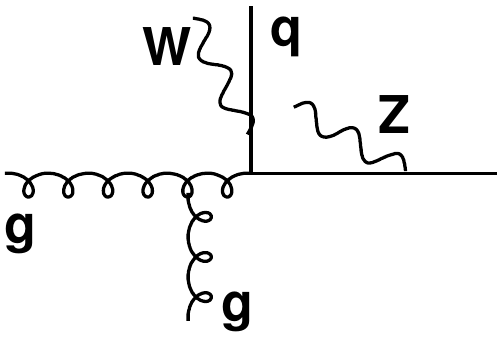}\\
 \end{minipage}
\begin{minipage}{1\columnwidth} 
\vskip 3pt
\hspace{1.3cm} LO\hspace{4.9cm} NLO\hspace{4.6cm} NNLO\\
 \end{minipage}
  \caption{\label{fig:diagram}%
  Example diagram contributing to WZ production at LO, NLO and NNLO.
  }

\end{figure}

For example, for $WZ$ production~(see Fig.~\ref{fig:diagram}), at LO only the $q\bar{q}\to
WZ$ production mode contributes, however, at NLO, new $gq~(g
\bar{q}) \to WZ q~(\bar{q})$ channel, with enhanced luminosity, appears. The corrections for $W\gamma$ production
can be even larger due to a radiation zero~\cite{Brown:1982xx} which suppresses
the LO contributions. Gluonic neutral radiation as part of the real corrections
breaks down the radiation pattern resulting in large NLO corrections.

The NLO differential distributions are sizable and phase space dependent --
therefore normalizing the LO differential distributions by the K-factor~(NLO/LO) of the
integrated cross section does not provide reliable predictions. Giant K-factors
of order 20 for commonly used observables can appear, which have a topological
phase space origin. At LO, e.g. for $WZ$ see Fig.~\ref{fig:diagram}~(left),
only back-to-back $WZ$ configurations are possible whereas at NLO, one
electroweak boson can recoil against one parton and the other can be emitted
arbitrarily soft or collinear, yielding large logarithmic enhancements.

Gluon-gluon-initiated contributions for neutral production processes with a
closed fermion loop, which are formally at QCD NNLO due to their one-loop
$\times$ one-loop nature, have also been computed in Refs.~\cite{gginduced}.
The corrections can be as large as 20$\%$, the $\alpha_s^2$ suppression is
compensated partially by the large gluonic pdfs. 

The QCD NLO corrections for all production modes, and also the NNLO QCD fermion-loop gluon initiated
contributions for neutral production channels, are available in the latest release
of the VBFNLO package~\cite{Arnold:2008rz}, which includes also anomalous coupling effects at NLO. 

Recently, EW corrections for almost all diboson production processes have
been computed in Refs.~\cite{dibosonEW} for on-shell production. The corrections can be sizable in the tails of the differential
distributions for commonly used experimental analyses. 

Because the NLO QCD corrections turn out to be extremely large, it is
important to assess the size of the NNLO QCD corrections. At this
order, as can be seen in Fig.~\ref{fig:diagram} for the WZ case, new topologies and
sub-processess appear first at NNLO, which can result in large
corrections. 

The full NNLO corrections for di-photon production have been provided in
Ref.~\cite{Catani:2011qz} and turned out to be sizable. The two-loop virtual corrections for
di-boson on-shell production have been presented in Refs.~\cite{virtualsNNLO}. Results for $VV +jet$ at NLO QCD, which
provide the one-loop real-virtual and double real corrections have been also computed
in a series of papers~\cite{dibosjet,Campanario:2010hp} and are available either in VBFNLO~\cite{Arnold:2008rz} or in MCFM~\cite{Campbell:2010ff}. 

Given the fact that $VV +$ jet at NLO QCD provides an essential piece of the
NNLO QCD corrections of $VV$ production, accounting  both for the new
sub-processes and topologies appearing first at NNLO, 
the question is whether we can use this information to provide approximate
results at NNLO for $VV$ production. %
%
%
We used the LOOPSIM method~\cite{Rubin:2010xp} to accomplish this and compute approximate NNLO QCD
corrections for $WZ$ production. %
The LOOPSIM method is based on unitarity and is able to merge processes with different jet multiplicities in a consistent
way. To produce approximate NNLO results for $WZ$
production~(\nNLO in LOOPSIM notation), the program needs to merge samples of $WZ$ and
$WZj$ at NLO accuracy. The $WZj$ samples at NLO, computed in
 Ref.~\cite{Campanario:2010hp}, are obtained from the VBFNLO package, which also provides the $WZ$ events at NLO. An interface
was created to communicate between the two programs for this purpose~\cite{Campanario:2012fk}. From the tree level and the one-loop correction 
events of $WZj$, LOOPSIM produces approximated 2-loop virtual counterterms for
$WZ$ production,  which are designed to cancel
the infrared divergences. %
The
LOOPSIM method has an internal parameter, $R_{LS}$, to evaluate the uncertainty. It will
be shown that this is  smaller than the remaining factorization and  renormalization scale uncertainties.

In the following, results at $\bar{n}\text{NLO}$ are given. They were studied
in Ref.~\cite{Campanario:2012fk}. The cuts applied were defined to closely
follow the experimental analyses,
\begin{equation}
p_{T,\ell}\ge 15(20), \quad |y_{l}|\le 2.5, \quad E_{T,\text{miss}} > 30
\GeV, \quad 60 < m_{l^+l^-} <
120 \GeV,
\end{equation}
where the parenthesis indicates cuts applied to leptons coming from Z.
For observables that involve jets, we
  consider only those jets that lie in the rapidity range
  $|y_{\text{jet}}|\le 4.5$ and have transverse momenta 
  $p_{T,\text{jet}}\ge 30\GeV$. The anti-$k_t$
  algorithm~\cite{Cacciari:2008gp} has been used, as implemented in FastJet~\cite{Cacciari:2005hq}, with the radius
$R=0.45$. Additionally, the leptons and jets are required to be well separated $\Delta R_{l(l,j)} > 0.3$.
For the central value of the renormalization and factorization scale we use
$\mu_0=\mu_R=\mu_F= (\sum
p_{T,partons}+\sqrt{p_{T,W}^2+M_W^2}+\sqrt{p_{T,Z}+M_Z^2})/2$. 

\begin{figure}[h!]
  \centering
  \includegraphics[width=0.45\columnwidth]{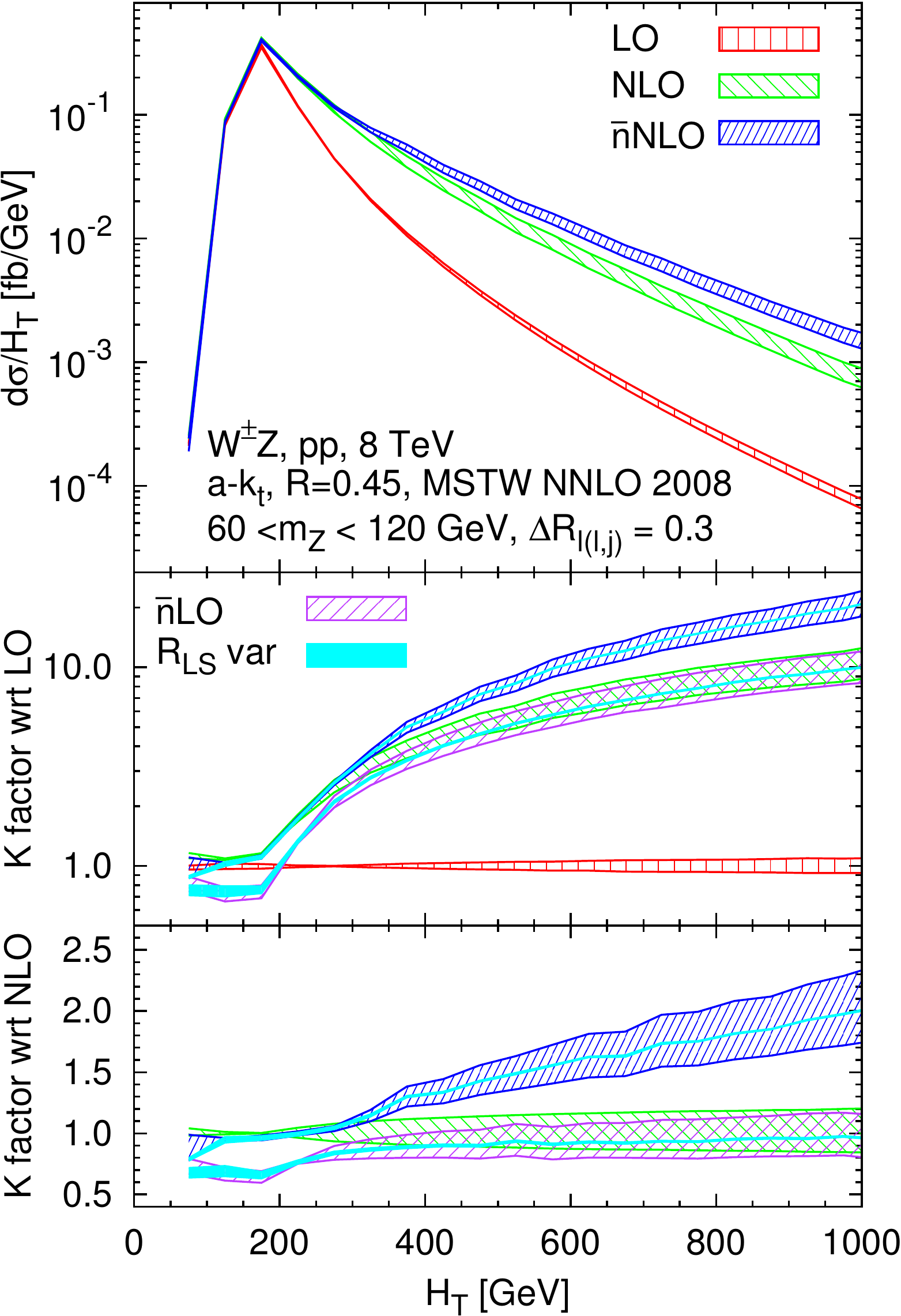}
  \hfill
  \includegraphics[width=0.45\columnwidth]{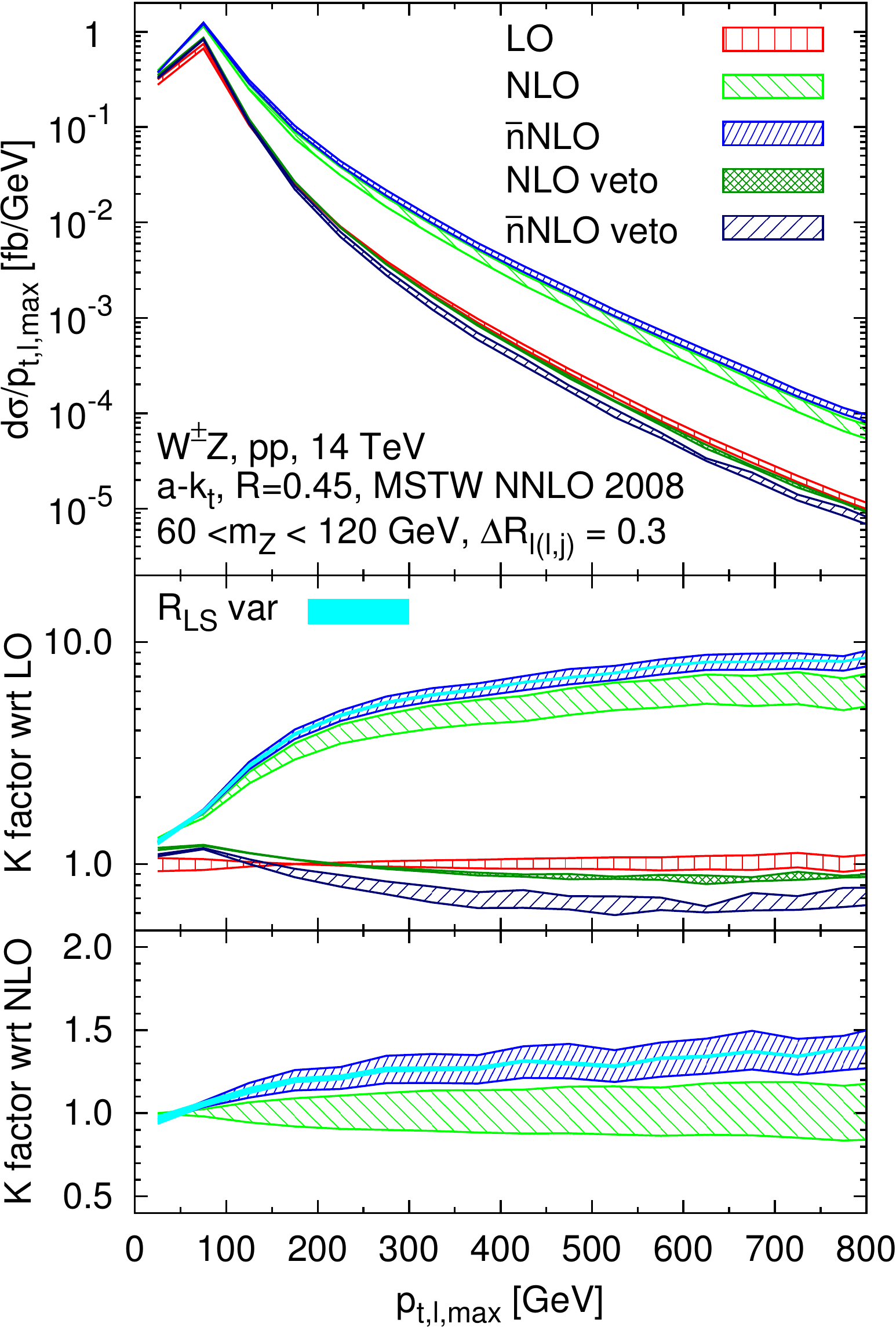}
  \caption{ \label{fig:pt_max}Differential cross sections and K factors for the effective mass observable,
  defined in Eq.~(\protect \ref{eq:HT}), for the LHC at $\sqrt{s}=8\,
  \text{TeV}$ (left). 
  Differential cross sections and K factors for the $p_T$ of the hardest lepton
  for the LHC at $\sqrt{s}=14\, \text{TeV}$
  (right). The bands correspond to varying $\mu_F=\mu_R$ by factors 1/2 and 2
  around the central value. The cyan solid bands give
  the uncertainty related to the $R_\text{LS}$ parameter varied between 0.5 and
  1.5.  The distribution are sums of contributions from two unlike flavor decay
  channels, $ee\mu\nu_{\mu}$ and $\mu\mu e\nu_e$.
  }
 \end{figure}

As a first check of our setup, we have merged WZ@LO and WZj@LO to
produce WZ@\nLO, which can be tested against the full WZ@NLO result. 
In the left panel of Fig.~\ref{fig:pt_max}, we show the effective mass defined by
\begin{equation}
H_{T} = \sum  p_{T,\text{jets}} + \sum  p_{T,l} +
E_{T,\text{miss}}\,.
\label{eq:HT}
\end{equation}

One can observe in the central panel that the \nLO result converges to the full
NLO result quickly, predicting correctly K-factors of order 10. The $\nNLO$
corrections can be as large as 100$\%$ compared to NLO (bottom panel of Fig.~\ref{fig:pt_max}) and they are clearly beyond the NLO scale uncertainties.
The R$_\text{LS}$ uncertainties are small in comparison to those from varying the factorization
and renormalization scales, which show a marginal reduction. The latter is
related to the fact that the $H_T$ observable favours regions of phase
space associated with new topologies entering first at NNLO, which are
computed only at LO. 

In the right panel of Fig.~\ref{fig:pt_max}, we present results for the differential distribution of the
lepton with higher transverse momenta. One observes that the \nNLO corrections
are large and beyond the scale uncertainty, reaching values as high as
40$\%$. The scale uncertainty is significantly improved and the $R_{LS}$
uncertainty is marginal. We include the vetoed sample to mimic some of the
experimental analyses. One can see that the \nNLO corrections are negative and
exhibit larger scale uncertainties than the NLO corrections, showing the known
feature of an artificially small scale uncertainty of NLO predictions with jet veto.

\section{Tri-boson production}
\label{sect:triboson}
Tri-boson production processes are important backgrounds to SM and BSM
searches.  They have been already examined in many experimental analyses. As a
signal, they allow us to quantify possible deviations from the Standard Model
coming from anomalous triple and quartic gauge couplings. From the theoretical
side, all the possible production modes have been computed at NLO QCD within
the VBFNLO collaboration in Refs.~\cite{tribos,Bozzi:2011wwa}, including the leptonic decays of the
vector bosons and all off-shell effects. Some of the channels have been
computed by other groups Refs.~\cite{tribosonshell,Baur:2010zf} for on-shell production
and neglecting Higgs boson exchange. The NLO QCD corrections are large ranging from 40$\%$- 100$\%$ at the
cross section level and even larger for differential distributions. The origin
of the size of the corrections is the same as for the di-boson case. At NLO,
new sub-processes and topologies appear.

\waa~production is the tri-boson production channel with the largest K-factor for
the integrated cross section ($\sim 3$) for standard
 experimental inclusive cuts. This feature, similarly to the $ W \gamma$ case, 
is due to the radiation zero~\cite{Brown:1982xx} pattern present at LO and broken at NLO by
 additional real QCD radiation, \waa~+jet, as
part of the NLO contributions. %
In Ref~\cite{Bozzi:2011wwa,Baur:2010zf}, it was shown that an additional 
jet veto-cut might help in the detection of the radiation
zero. However, due to the aforementioned
problem with the exclusive vetoed samples, this procedure raises the
question of the reliability of the predictions. Additionally, the remaining
scale uncertainties at NLO QCD are due to unbalanced
gluon-induced real radiation computed at LO, e.g., $gq\to$ \waa~$q$. 
Thus, to realistically assess the uncertainties, also concerning anomalous coupling searches and as an
important step towards a NNLO QCD or a LOOPSIM \nNLO QCD calculation of \waa~,  we calculated \waa~+jet at NLO QCD.  


\waa + jet at NLO QCD includes the
evaluation of the complex hexagon virtual amplitudes computed in
Ref.~\cite{Campanario:2011cs}. These pose a
challenge not only at the level of the analytical calculation, but also
concerning the CPU time required to perform a full $2\to 4$ process at
NLO QCD. 

\begin{figure*}[h!]
 \includegraphics[width=0.45\columnwidth]{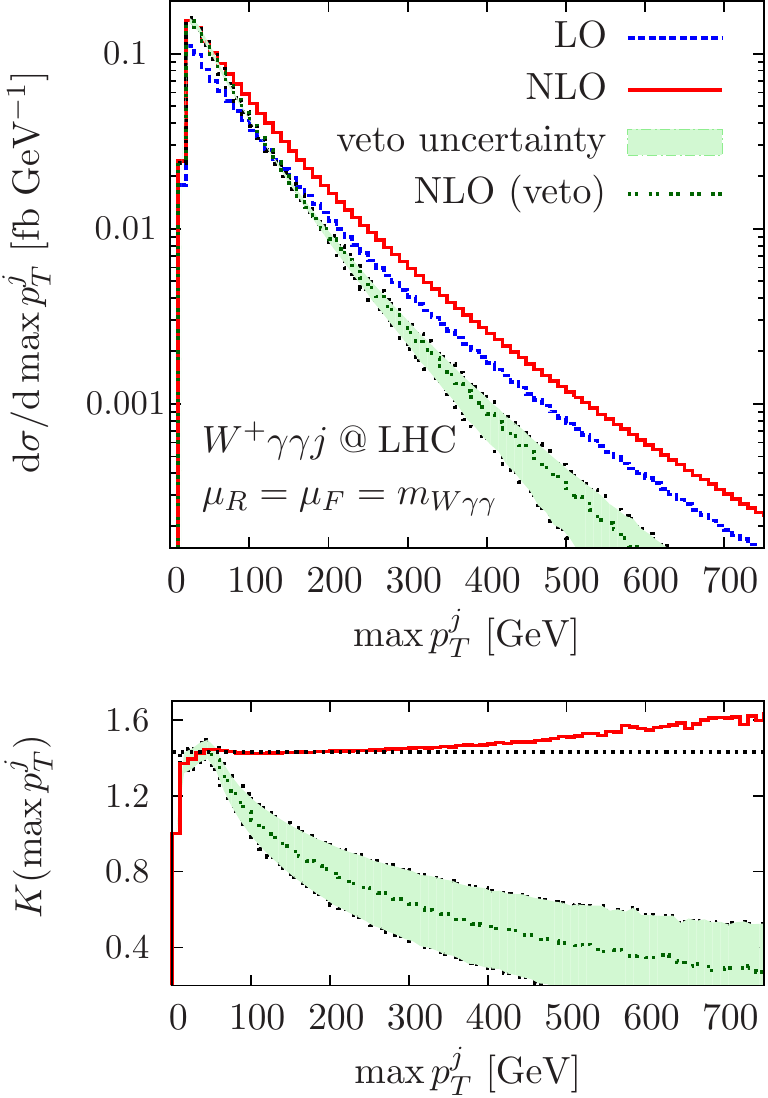}
\hspace*{1cm}
 \includegraphics[width=0.45\columnwidth]{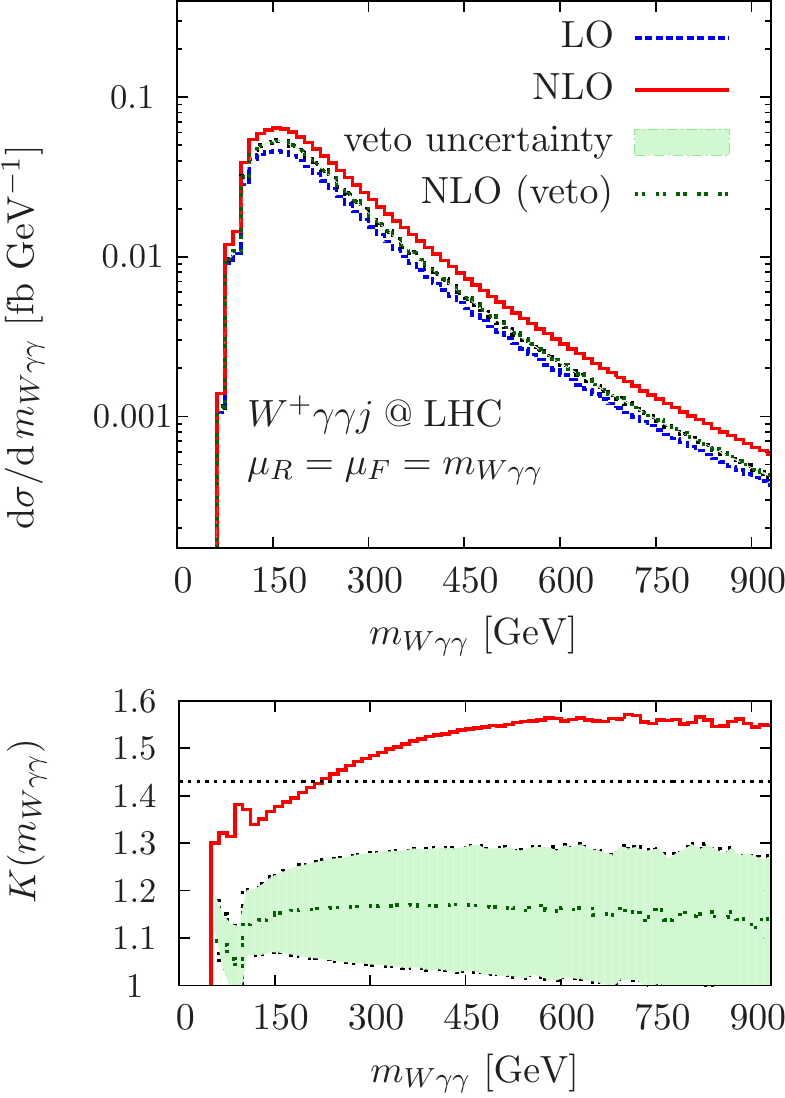}
\caption{\label{Camp:fig:waaj:diff} Differential $\max p_T^j$ and
      $m_{W\gamma\gamma}$ distribution for inclusive and exclusive
      $l^+\bar \nu \gamma\gamma$+jet production.}
\end{figure*}

The details on the setup can be found in 
Ref.~\cite{Campanario:2011ud}.  We consider $W^\pm$ decays to 
the first two lepton generations, i.e, $W\rightarrow
e\nu_e(+\gamma),\mu\nu_\mu(+\gamma)$. Both contributions are summed in Fig~\ref{Camp:fig:waaj:diff}.

K-factors of about $1.4$, which are similar to the ones found in other multi-boson+jet
processes~\cite{dibosjet,Campanario:2010hp} are found 
 for the LHC at $\sqrt{\text{s}} =14 $ TeV. This moderate K-factor as compared
  to corrections of  $\sim 300\%$ for \waa~production confirms that the large
  integrated K-factor encountered in \waa~ production can mainly be attributed
  to radiation zero cancellations, which are not present in \waa~+ jet.  The scale dependence of the
  $W\gamma\gamma j$ production cross section turns out to be modest at about
  10$\%$. The phase space dependence of the QCD corrections is
  non-trivial and sizable.  Additional parton emission modifies
the transverse momentum and invariant mass spectra. The leading jet becomes
slightly harder at NLO as can be inferred from the differential $K$
factor in the bottom panel of Fig.~\ref{Camp:fig:waaj:diff}. Vetoed
real-emission distributions are plagued with large
uncertainties (Fig.~\ref{Camp:fig:waaj:diff}, left). 

\section{Summary}
Results beyond NLQ QCD have been presented for $WZ$ production at the LHC
as well as for \waa+ jet at NLO. The corrections are beyond scale
uncertainties and exhibit a non-trivial phase space dependence. When comparing
precisely measured distributions in these channels against Monte Carlo
predictions, un-included QCD corrections could be misinterpreted
for anomalous electroweak trilinear or quartic couplings arising from
new interactions beyond the SM.
\section*{Acknowledgments} 
We acknowledge support by the Deutsche Forschungsgemeinschaft under SFB TR-9 ``Computergest\"utzte
Theoretische Teilchenphysik''. 
FC is funded by a Marie Curie fellowship (PIEF-GA-2011-298960) and partially by
MINECO (FPA2011-23596) and by LHCPhenonet (PITN-GA-2010-264564).  
MR acknowledges support by the Initiative and Networking Fund of the
Helmholtz Association, contract HA-101(``Physics at the Terascale''). 
SS was in part supported by European Commission under contract
PITN-GA-2010-264564.

\end{document}